# Physical meaning of vorticity based on the RS decomposition and explicit formula for the Liutex vector

Yiqian Wang[1](王义乾) , Yisheng Gao[2] (高宜胜), Chaoqun Liu[2*]

1. School of Aerospace Engineering, Tsinghua University, Beijing 100084, China
2. Department of Mathematics, University of Texas at Arlington, Arlington 76019, USA

**Abstract:** In the present study, the physical meaning of vorticity is revisited based on the RS decomposition proposed by Liu *et al.* in the framework of Liutex (previously named Rortex), a vortex vector field with information of both rotation axis and swirling strength [C. Liu *et al.*, "Rortex—A new vortex vector definition and vorticity tensor and vector decompositions", Phys. Fluids 30, 035103 (2018)]. It is demonstrated that the vorticity in the direction of rotational axis is twice the spatial mean angular velocity in the small neighborhood around the considered point while the imaginary part of the complex eigenvalue ($\lambda_{ci}$) of the velocity gradient tensor (if exist) is the pseudo-time average angular velocity of a trajectory moving circularly or spirally around the axis. In addition, an explicit expression of the Liutex vector in terms of the eigenvalues and eigenvectors of velocity gradient is obtained for the first time from above understanding, which can further, though mildly, accelerate the calculation and give more physical comprehension of the Liutex vector.

## Nomenclature:

| | |
|---|---|
| $\vec{v}$ | = components of velocity vector in the original $xyz$-frame |
| $\vec{V}_Q$ | = components of velocity vector in $XYZ_Q$-frame |
| $\vec{V}_P$ | = components of velocity vector in $XYZ_P$-frame |
| $Q$ | = rotation matrix to rotate from xyz-frame to $XYZ_Q$-frame |
| $P$ | = rotation matrix to rotate from $XYZ_Q$-frame to $XYZ_P$-frame |
| $u, v, w$ | = velocity components in $xyz$-frame |
| $U_Q, V_Q, W_Q$ | = velocity components in $XYZ_Q$-frame |
| $U_P, V_P, W_P$ | = velocity components in $XYZ_P$-frame |
| $\nabla$ | = the operator to get the gradient tensor of a vector |
| $\vec{R}, R$ | = The Liutex vector and its magnitude |

**Corresponding author:** Chaoqun Liu, Email: cliu@uta.edu

| | |
|---|---|
| $\theta$ | = The rotation angle of the $P$ rotation |
| $\alpha, \beta$ | = Intermediate parameters to determine the magnitude of Liutex vector |
| $\lambda_{cr} \pm i\lambda_{ci}, \lambda_r$ | = The complex conjugate and real eigenvalues of velocity gradient tensor |
| $\vec{r}$ | = The real eigenvector of velocity gradient tensor corresponding to $\lambda_r$ |
| $d\vec{s}$ | = The vector pointing from a point A to its surrounding point B |
| $T, T^{-1}$ | = The eigenvector formed matrix and its inverse |
| $\dot{\theta}$ | = The angular velocity of the fluid particles rotating around the axis |
| $\dot{\theta}_{max}, \dot{\theta}_{min}$ | = Then maximum and minimum $\dot{\theta}$ |
| $\omega_x, \omega_y, \omega_z$ | = Vorticity components of the vorticity vector |
| $T_p$ | = The pseudo-time period of the fluid particles rotating around the axis |

# 1. Introduction

Vorticity is rigorously and mathematically defined as the curl of the velocity field. However, its physical meaning remains ambiguous even after many year intense study. It is very commonly found in textbooks that vorticity can be interpreted as twice of the angular velocity of a fluid element. As stated by Wu et al.[1], however, the concept of angular velocity is originally defined for rigid motion, which is not the case for deformable fluid[2,3]. As a counter example, there is concentration of vorticity near the wall surface in a Blasius boundary layer undoubtedly. However, the streamlines and pathlines are basically straight lines which is certainly not the case of vorticity in rigid motion. To remedy this, the concept of angular velocity could be extended as how one fluid particle rotates to a certain center and thus the isotropy in rigid motion is lost in fluids. Consequently, vorticity can be viewed as a measure of local spinning motion of the fluid near some point and could be quantified as twice the mean angular velocity vector of those particles relative to their center of mass. Most discussion of vorticity in literature ends at this point. However, we emphasize here vorticity is actually the spatial mean angular velocity that has a counterpart of pseudo-time mean angular velocity which will be discussed later.

In any manner, vorticity is related to the rotational motion of a fluid such that a flow field is called irrotational if the vorticity is zero everywhere, otherwise, the flow is called rotational. Naturally, vorticity became the first quantity people resort to detect vortices in the flow, which has clear physical intuition as a fluid region with rotational motions but no mathematical definition until the recent introduction of Liutex vector (previously named Rortex). The concept of vortex is especially important in turbulent flows as its close connection to coherent structures, *i.e.*, organized flow structures in turbulence which play a primary role in turbulent momentum transport[4,5]. Hairpin vortices and near-wall low-speed streaks are two most representative coherent structures. Back in 1952, Theordorsen[6] proposed a conceptual "horseshoe" or "hairpin" vortex model to describe the generation and sustenance of turbulence. These hairpin shaped vortices are commonly found in turbulence both from experiments and numerical simulations. Adrian[7] addressed hairpins can autogenerate to form hairpin packets which are the prevalent coherent structures in wall turbulence. The development from Λ vortex to hairpin vortex in boundary layer transition was carefully studied by Wang et al.[8] and the preponderance and statistical importance of hairpins during transition and in turbulence was also further explored[9]. The second coherent structure discussed here is the near-wall streaks of alternating low and high streamwise momentum fluids which were first reported by Kline et al.[10] in 1967 from flow visualization using hydrogen bubbles in the viscous sublayer. Since then the streaks and surrounding staggered quasi-streamwise vortices received substantial attention and the dynamics of a self-sustained nonlinear cycle involving the instability of streaks leading to the formation of longitudinal vortices is proposed[11]. Obviously, both hairpins and streaks are related to vortices, a concept with clear physical intuition but hardly a mathematical definition. Eulerian velocity-gradient-based scalar vortex identification methods including $\Delta$, $Q$, $\lambda_{ci}$ and $\Omega$ method[12–15] have been proved to be able to capture the rotational strength of the vortices to some extent. In addition, the effect of compressibility and the presence of wall surface on vortex identification are further explored[16,17], which clearly show that the vortical structures identified by these methods could be seriously contaminated by shearing under circumstances. Reviews of vortex identification methods can be found in the literature by Zhang et al.[18] and Epps[19].

To give a more precise and unambiguous definition of a vortex, a vector field which includes information of both direction and magnitude named Liutex[20,21] (previously named Rortex) was introduced recently. Gao and Liu[22] substantially improved the method by pointing out that the local rotational axis is the real eigenvector of the velocity gradient tensor given that the other two eigenvalues are complex conjugates which is the necessary and sufficient condition that the local fluid is rotating. Then, the swirling strength is determined in the plane perpendicular to the rotational axis, which could be interpreted as the rotational part of vorticity $R$ with a residual vorticity part $S$. According to this RS decomposition, the vorticity vector could be further decomposed into a rotational part R, which represents the rigid-body rotation part and a non-rotational part S, which is mainly shear. The computational procedure of Liutex vector includes two-step coordinate rotation of $Q$ rotation and $P$ rotation. In $Q$ rotation, the z-axis is rotated to the direction of rotational axis, which is actually the real eigenvector of velocity gradient tensor. The second $P$ rotation aims to get the minimum angular velocity about the rotational axis, which is then chosen as the magnitude of Liutex vector. Despite the sound deduction of the procedure, an explicit formula of the Liutex vector in terms of quantities in the original reference frame is nevertheless missing. The issue will be dealt with and for the first time such an explicit expression of Liutex vector will be given in this paper.

Following the background introduction to the concept of vorticity and vortex in this section, the mathematical definition of a vortex *i.e.*, Liutex vector and its calculation procedure is revisited in Section 2. In Section 3, the trajectories of surrounding fluid particles of a point A based on the velocity gradient is carefully analyzed. It is concluded that the vorticity along the direction of rotational axis is twice the spatial mean angular velocity about the rotational axis, $\lambda_{ci}$ (the imaginary part of the complex eigenvalue of velocity gradient tensor) is the pseudo-time average angular velocity, while the magnitude of Liutex vector is twice the minimum angular velocity. Based on these physical understanding of the three quantities, an analysis of which quantity can best describe the rotational motion of the fluid is given and the selection of the minimum angular velocity in the frame of Liutex vector is thus justified with the physical understanding of decomposing vorticity into a rotational part $R$ and a non-rotational part $S$. An explicit expression of Liutex vector in terms of eigenvalues and eigenvectors of velocity gradient tensor is then introduced for the first time in Section 4 with discussion of its physical meaning and computational efficiency improvement. In Section 5, we demonstrate the Liutex vector method and RS decomposition as a powerful tool to analyze the flow field by examples of a boundary layer transition and a turbulent channel flow. Then the paper ends with conclusions in Section 6.

## 2. Definition of Liutex vector and RS decomposition

The improved computational procedure of Liutex vector consists of the following steps:

1) Get the direction of Liutex

Evaluate the eigenvalues of the velocity gradient tensor $\nabla \vec{v}$ in the original $xyz$-frame. If all three eigenvalues are real, no fluid rotation exists, thus the Liutex vector is set to zero; if there are two complex conjugate and one real eigenvalues, calculate the corresponding unit real eigenvector $\vec{r}$, which represents the local rotational axis and thus the direction of Liutex. And then, make a coordinate rotation ($Q$ rotation) which rotates the original $z$-axis to the direction of the local rotational axis $\vec{r}$ and calculate the velocity gradient tensor $\nabla \vec{V}_Q$ in the resulting $XYZ_Q$ frame by $\nabla \vec{V}_Q = Q \nabla \vec{v} Q^\mathrm{T}$.

2) Obtain the rotational strength (magnitude) of Liutex

After the $Q$ rotation, the velocity gradient tensor $\nabla \vec{V}_Q$ has the form of

$$\nabla \vec{V}_Q = \begin{bmatrix} \frac{\partial U_Q}{\partial X_Q} & \frac{\partial U_Q}{\partial Y_Q} & 0 \\ \frac{\partial V_Q}{\partial X_Q} & \frac{\partial V_Q}{\partial Y_Q} & 0 \\ \frac{\partial W_Q}{\partial X_Q} & \frac{\partial W_Q}{\partial Y_Q} & \frac{\partial W_Q}{\partial Z_Q} \end{bmatrix} \tag{1}$$

A second rotation ($P$ rotation) is used to rotate the reference frame around the $Z_Q$-axis and the corresponding velocity gradient tensor $\nabla \vec{V}_P$ can be written as

$$\nabla \vec{V}_P = P \nabla \vec{V}_Q P^{-1} \tag{2}$$

with

$$P = \begin{bmatrix} \cos\theta & \sin\theta & 0 \\ -\sin\theta & \cos\theta & 0 \\ 0 & 0 & 1 \end{bmatrix} \tag{3}$$

The new $\partial U_P / \partial Y_P$ with rotation angle of $\theta$ can be regarded as the angular velocity of fluid at this azimuth angle $\theta$, and can be written as

$$\frac{\partial U_P}{\partial Y_P}\bigg|_\theta = \alpha \sin(2\theta + \varphi) - \beta \tag{4}$$

where $\alpha$ and $\beta$ are determined by elements of

$$\alpha = \frac{1}{2}\sqrt{\left(\frac{\partial V_Q}{\partial Y_Q} - \frac{\partial U_Q}{\partial X_Q}\right)^2 + \left(\frac{\partial V_Q}{\partial X_Q} + \frac{\partial U_Q}{\partial Y_Q}\right)^2} \tag{5}$$

$$\beta = \frac{1}{2}\left(\frac{\partial V_Q}{\partial X_Q} - \frac{\partial U_Q}{\partial Y_Q}\right) \tag{6}$$

and $\varphi$ is a constant angle determined by $\nabla \vec{V}_Q$. Then the rotational strength is defined as twice the minimal absolute value of the off-diagonal component of the $2 \times 2$ upper left submatrix, i.e., the minimal absolute value of $\partial U_P / \partial Y_P |_\theta$ and can be given by

$$R = \begin{cases} 2(\beta - \alpha), & \alpha^2 - \beta^2 < 0 \\ 0, & \alpha^2 - \beta^2 \geq 0 \end{cases} \quad (7)$$

and here we assume $\beta > 0$ (if $\beta < 0$, we can first rotate the local rotation axis to the opposite direction to make $\beta$ positive).

Finally, the Liutex vector is obtained as $\vec{R} = R\vec{r}$.

## 3. Analysis on the velocity gradient tensor

A. Problem formulation

Suppose a point $A$ in the flow field with gradient $\nabla \vec{v}_A$. Then it's neighborhood region can be defined as a region with distance from point A satisfying $L < \delta$, where $\delta$ is an arbitrary small quantity. Given that $\delta$ is small enough, the velocity of a point $B$ inside this neighbourhood can be approximated as:

$$\vec{v}_B = \vec{v}_A + \nabla \vec{v}_A d\vec{s} \quad (8)$$

where $\vec{v}_A$ and $\vec{v}_B$ are the velocity vector at point $A$ and $B$ respectively while $d\vec{s}$ is the vector pointing from $A$ to $B$. Since we only care how the surrounding fluids move about point $A$, without loss of generality we assume $\vec{v}_A = 0$ and put the origin of the coordinates at point $A$. Thus, $d\vec{s}$ can be expressed as the coordinates of point $B$, i.e., $d\vec{s} = [x_B \ y_B \ z_B]^T$. In addition, $\frac{d}{dt}[x_B \ y_B \ z_B]^T = \vec{v}_B$, which leads to

$$\frac{d}{dt}\begin{bmatrix} x_B \\ y_B \\ z_B \end{bmatrix} = \nabla \vec{v}_A \begin{bmatrix} x_B \\ y_B \\ z_B \end{bmatrix} \quad (9)$$

The above equation describes the movement of surrounding fluid particles in a small enough neighbourhood (represented by point $B$) relative to point $A$ with velocity gradient $\nabla \vec{v}_A$. For this linear equation to be valid, (1) point $B$ has to be sufficient close to $A$ while here we also assume the continuity hypothesis to be held; (2) the time span considered is sufficient small. As a first-order linear system, the solution to Equation 9 heavily depends on the eigenvalues and eigenvector of $\nabla \vec{v}_A$. There are two scenarios regarding the eigenvalues of a 3 by 3 matrix: (1) three real eigenvalues;

(2) one real and two complex conjugate eigenvalues. For the first scenario, the fluid is either stretched or compressed in the directions of three real eigenvectors. On the other hand, the fluid undergoes rotational motion due to the existence of complex conjugate eigenvalues in the second situation. Notice that in both $\lambda_{ci}$ method and the Liutex method, the condition that whether complex eigenvalues of velocity gradient tensor exist is used to determine whether a vortex region is detected. However, Gao and Liu[22] further pointed out that the rotational axis is actually the real eigenvector, along which only stretch or compression can be found. Since we are more concerned about vortex related phenomenon, mainly the second scenario is discussed here. Therefore, the solution to equation 9 has the form of:

$$\begin{bmatrix} x \\ y \\ z \end{bmatrix} = T \begin{bmatrix} e^{(\lambda_{cr}+i\lambda_{ci})t} & & \\ & e^{(\lambda_{cr}-i\lambda_{ci})t} & \\ & & e^{\lambda_r t} \end{bmatrix} T^{-1} \begin{bmatrix} x_0 \\ y_0 \\ z_0 \end{bmatrix} \quad (10)$$

where $\lambda_{1,2} = \lambda_{cr} \pm i\lambda_{ci}$ and $\lambda_r$ are the complex conjugate and real eigenvalues respectively while $T$ is a matrix formed by the corresponding eigenvectors with $T^{-1}$ as its inverse. $[x_0 \; y_0 \; z_0]^T$ is the initial position of the point B while $[x \; y \; z]^T$ is the solution to the system. Notice the subscript of B is left out for simplicity from now on.

B. Analysis on a point from boundary layer transition

Here, we select a point near the central plane of a hairpin vortex in a natural boundary layer transition as point $A$ with velocity gradient tensor

$$\nabla \vec{v} = \begin{bmatrix} 0.0760 & 0.0520 & 0.3215 \\ -0.0134 & 0.0101 & 0.0009 \\ -0.1342 & -0.0188 & -0.0915 \end{bmatrix} \quad (11)$$

whose eigenvalues are $\lambda_{1,2} = \lambda_{cr} \pm i\lambda_{ci} = -0.0018 \pm 0.1920i$ and $\lambda_r = 0.0103$. Based on Equation 10, the solution can be visualized as shown in Figure 1 (a) with the filled magenta circle being point A, the straight red line being the real eigenvector, and the remaining four curves being pathlines originate from corresponding seeding points or initial conditions denoted as blue squares. It should be noted that since the system is linear, the pathlines are identical to streamlines. In addition, despite the coordinate range presented in Figure 1 (a), the coordinates can always multiply an arbitrary small number in order to fulfill the assumption point $B$ is sufficiently close to point $A$ to be valid. On the other hand, the time-span in which the real trajectories will follow the curves in Figure 1 (a) could be small so that it cannot even make a full circular or spiral motion. However, the trajectories could still be meaningful and give information of the short-time behavior of the surrounding fluid particles. Obviously from Figure 1 (a), the surrounding fluid is making spiral movement around the real eigenvector, which can be seen clearer in Figure 1 (b) with the observing plane being normal to the real eigenvector. Another observation is that the trajectory is generally elliptic as shown in Figure 1 (b).

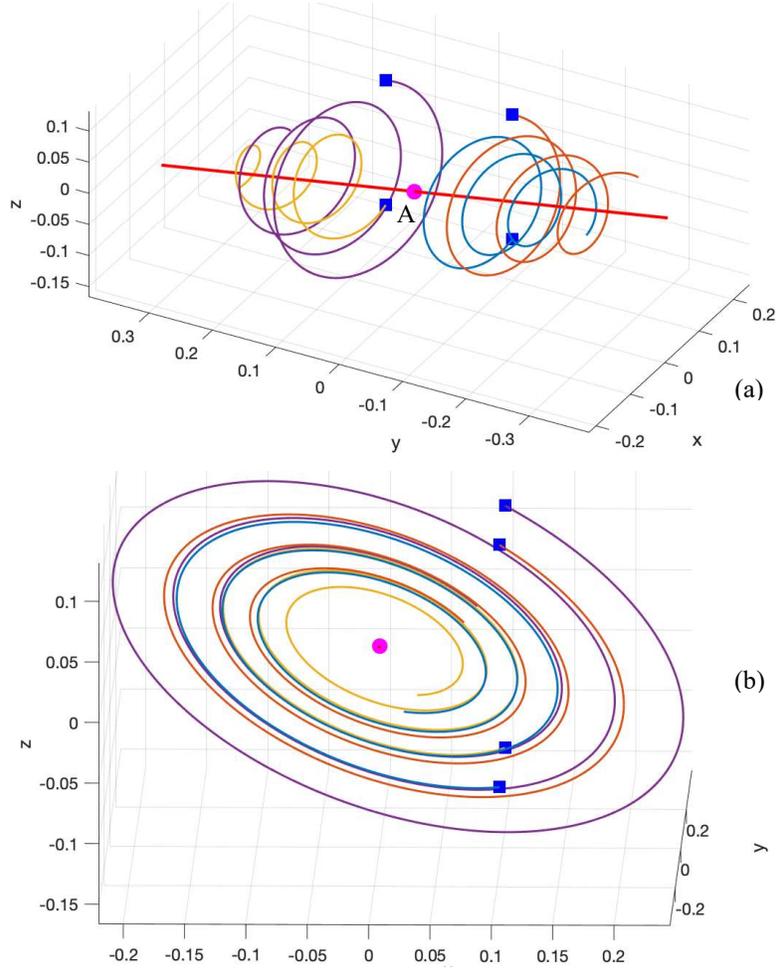

Figure 1. The fluid motion around a point A in boundary layer transition.

In the framework of Liutex, the $Q$ rotation rotates the original z-axis to the direction of the local rotational axis, *i.e.*, the real eigenvector $\vec{r}$. For the point A considered, the velocity gradient tensor after $Q$ rotation becomes

$$\nabla \vec{V}_Q = \begin{bmatrix} 0.0728 & 0.3288 & 0 \\ -0.1318 & -0.0884 & 0 \\ -0.0061 & -0.0104 & 0.0103 \end{bmatrix} \quad (12)$$

According to Equation 10, the new trajectories are shown in Figure 2 which are the same as the trajectories in Figure 1 except the orientation of the real eigenvector now points to the positive $Z_Q$ axis.

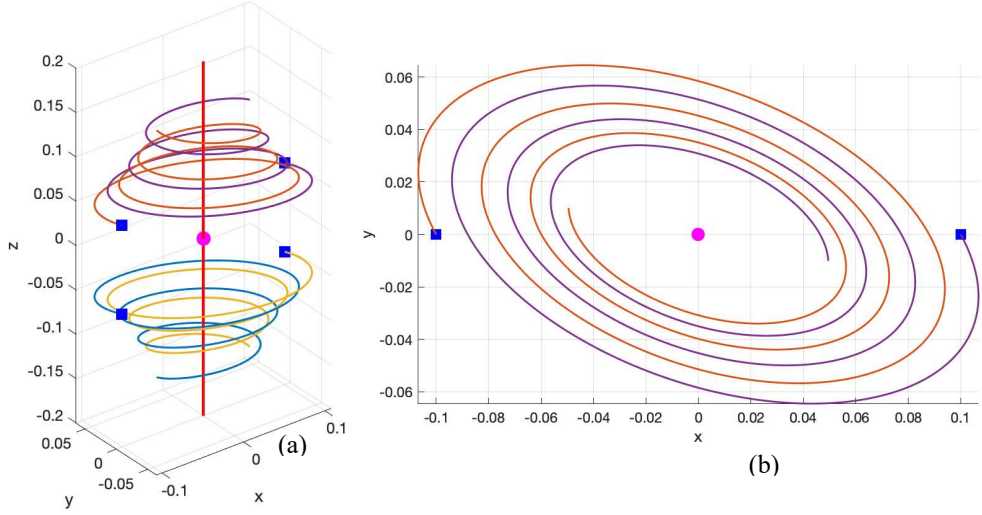

Figure 2. The fluid motion around point A after $Q$ rotation.

A second $P$ rotation is then applied to rotate the reference frame such that the angular velocity of the newer $Y_P$-axis, i.e., $\partial U_P/\partial Y_P$ reaches its minimum. The velocity gradient tensor now becomes

$$\nabla \vec{V}_P = \begin{bmatrix} -0.0078 & 0.1030 & 0 \\ -0.3576 & -0.0078 & 0 \\ -0.0118 & -0.0022 & 0.0103 \end{bmatrix} \quad (13)$$

The newer trajectories are shown in Figure 3. It is shown that besides the fact that the rotational axis is along the z-axis, the major axis and the minor axis of the elliptic shape formed by the trajectories coincide with the new $Y_P$ and $X_P$-axes respectively. $\partial U_P/\partial Y_P = 0.1030$ is the minimum angular velocity of $Y_P$-axis and thus $2 \times \partial U_P/\partial Y_P = 0.2060$ is the magnitude of Liutex vector.

According to Equation 4, the angular velocity of the newer $Y_P$-axis after $P$ rotation is a sine function of $2\theta$, which is defined as the in-plane rotation angle about the $Z_Q$-axis after $Q$ rotation. Here we define the angle $\theta$ backwardly, i.e., $\theta$ is the rotation angle from the optimum reference frame by a $P$ rotation as shown in Figure 3 (b). Thus, the angular velocity around point $A$ can be expressed as

$$\dot{\theta} = \frac{(\dot{\theta}_{max} - \dot{\theta}_{min})}{2} \sin\left(2\theta - \frac{\pi}{2}\right) + \frac{\dot{\theta}_{max} + \dot{\theta}_{min}}{2} \quad (14)$$

Clearly, $\dot{\theta}$ reaches its minimum of $\dot{\theta}_{min} = \partial U_P/\partial Y_P = 0.1030$ with $\theta = 0$ as P1 in Figure 3 (b), and reaches its maximum of $\dot{\theta}_{max} = -\partial V_P/\partial X_P = 0.3576$ with $\theta = \pi/2$ as P2 in Figure 3(b). The variation of $\dot{\theta}$ with $\theta$ around point $A$ is shown in Figure 4 with indication of point P1 and P2.

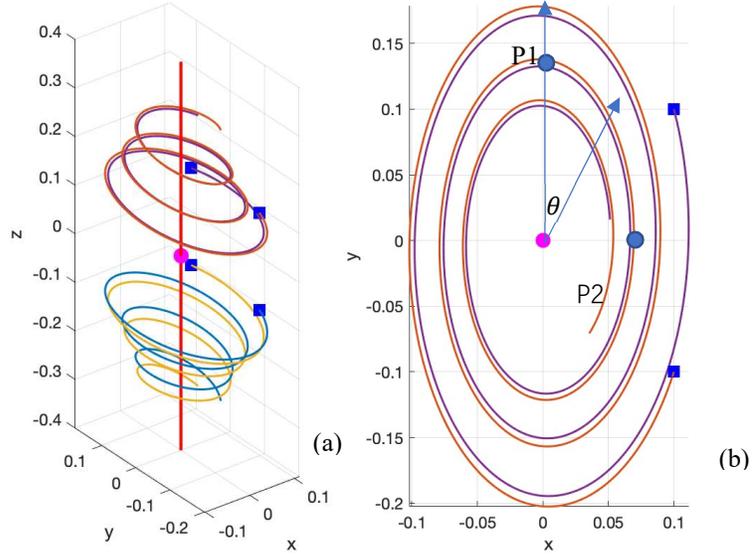

Figure 3. The fluid motion around point A after $P$ rotation.

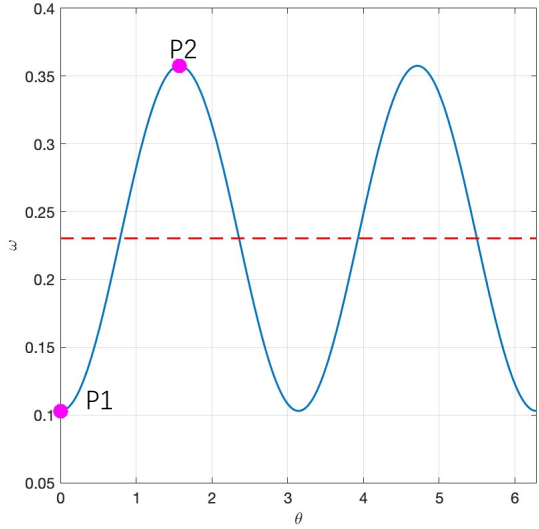

Figure 4. The angular velocity $\dot{\theta}$ as a function of $\theta$

Thus, the mean angular velocity $\langle \dot{\theta} \rangle$ along a trajectory in space (in $\theta$) is

$$\langle \dot{\theta} \rangle = \frac{1}{2\pi}\int_0^{2\pi} \dot{\theta}\, d\theta = \frac{\dot{\theta}_{max} + \dot{\theta}_{min}}{2} = \frac{\partial U_P/\partial Y_P - \partial V_P/\partial X_P}{2} = \frac{1}{2}\omega_{Z_P} \qquad (15)$$

where the subscript $P$ denotes the value is evaluated after the $P$ rotation, $\omega_{Z_P}$ is the vorticity in the $Z_P$ direction, and thus $\omega_{Z_P}$ is actually the vorticity in the direction of real eigenvector, i.e., $\omega_{Z_P} = \langle \vec{\omega}, \vec{r} \rangle$. Here, $\langle \vec{\omega}, \vec{r} \rangle$ denotes the dot product of $\vec{\omega}$ and $\vec{r}$. It should be noted that the unit real eigenvector is selected so that $\langle \vec{\omega}, \vec{r} \rangle$ is positive. Therefore, we can see that the spatial average angular velocity is actually equals half the vorticity magnitude along the real eigenvector. With reference to Equation 10, the vorticity components of $\omega_{X_P}$ and $\omega_{Y_P}$ represent the modulation effect of the circular or spiral motion in the direction of the rotational axis in addition to the stretch or compression determined by the real eigenvalue of the velocity gradient tensor, which are not as significant to the rotational motion compared with $\omega_{Z_P}$.

On the other hand, according to Equation 10 the motion of the surrounding particles of point $A$ has a "period" of $T_p = 2\pi/\lambda_{ci}$ if we ignore the stretch or compression represented by the real eigenvalue and the movement towards or away from the rotation axis determined by the real part of the complex eigenvalue, i.e., $\lambda_{cr}$. Under these assumptions, the trajectory can form a closed curve which make the motion periodic. The statement above is equivalent to that in a time-span of $T_p = 2\pi/\lambda_{ci}$, the surrounding fluid particle can rotate an angle of $2\pi$. Thus, the time-average angular velocity is $2\pi/T_p = \lambda_{ci}$. As stated in the previous section, for Equation 9 to be valid, the time-span considered must be small enough so that the fluid maybe not able to make a full circular motion. However, for this linear system, the pathlines and the streamlines are identical. Thus, the time-averaged angular velocity still has an implication of the instantaneous flow field. To be more precisely, we'll denote $\lambda_{ci}$ as the pseudo time-average angular velocity.

C. Discussion on the physical meaning of vorticity and $\lambda_{ci}$

From above discussion, one possible understanding of the vorticity along the direction of real eigenvector is twice the spatial mean angular velocity while $\lambda_{ci}$ can be viewed as the pseudo time-average angular velocity. Clearly, there must be some relation between these two quantities. First, from Equation 15 we can get

$$\frac{1}{2}\omega_{Z_P} = \frac{1}{2\pi}\int_0^{2\pi} \dot{\theta}\, d\theta = \frac{\dot{\theta}_{max}+\dot{\theta}_{min}}{2} \tag{16}$$

The time-period of $T_p$ can also be related to $\dot{\theta}$ as

$$T_p = \int_0^{2\pi} \frac{d\theta}{\dot{\theta}} \tag{17}$$

with

$$\dot{\theta} = \frac{(\dot{\theta}_{max}-\dot{\theta}_{min})}{2}\sin\left(2\theta - \frac{\pi}{2}\right) + \frac{\dot{\theta}_{max}+\dot{\theta}_{min}}{2} \tag{18}$$

The integral in Equation can be solved and finally

$$\lambda_{ci} = \frac{2\pi}{T_p} = \sqrt{\dot{\theta}_{min}\dot{\theta}_{max}} \tag{19}$$

The above equation can also be verified be checking the elements of $\nabla\vec{V}_P$ as

$$\nabla \vec{V}_P = \begin{bmatrix} \lambda_{cr} & \dot{\theta}_{min} & 0 \\ -\dot{\theta}_{max} & \lambda_{cr} & 0 \\ -0.0118 & -0.0022 & \lambda_r \end{bmatrix} \quad (20)$$

Its characteristic equation is

$$(\lambda - \lambda_r)[(\lambda - \lambda_r)^2 + \dot{\theta}_{min}\dot{\theta}_{max}] = 0 \quad (21)$$

To make its eigenvalues remain $\lambda_{1,2} = \lambda_{cr} \pm i\lambda_{ci}$ and $\lambda_r$, $\dot{\theta}_{min}\dot{\theta}_{max}$ has to equal to $\lambda_{ci}^2$.

In the frame work of Liutex, the magnitude is defined as twice the minimum angular velocity, *i.e.*,

$$R = 2\dot{\theta}_{min}$$

Thus, these three quantities, the vorticity in the direction of real eigenvector $\omega_{Z_P}$, the imaginary part of the complex eigenvalue $\lambda_{ci}$ and the magnitude of Liutex $R$ are all connected to the minimum and maximum angular velocity in the plane normal to the real eigenvector. In addition, these three quantities are clearly related to the rotational motion of the fluids around the considered point $A$ since their physical meaning stated above. However, it remains a question to be answered that which quantity is more appropriate to describe the swirling strength.

D. Two-dimension analysis

In this section, we focus our attention to the rotational motion in the plane normal to the real eigenvector by ignoring the third dimension after the $P$ rotation. In addition, $\lambda_{cr}$ represents the movement towards or away from the rotational axis which we also ignore here. The result two-dimensional (2D) velocity gradient tensor of point A becomes

$$\nabla \vec{V}_{2D} = \begin{bmatrix} 0 & \dot{\theta}_{min} \\ -\dot{\theta}_{max} & 0 \end{bmatrix} = \begin{bmatrix} 0 & 0.1030 \\ -0.3576 & 0 \end{bmatrix} \quad (22)$$

The typical induced pathline or streamline is shown in Figure 5 (a) as the blue curve, with corresponding angular velocity $\dot{\theta}$ as function of $\theta$ shown in Figure 5 (b) as blue curve. With vorticity fixed, i.e., the spatial mean angular velocity fixed shown as black dashed line in Figure 5 (b), we can change $\nabla \vec{V}_{2D}$ to

$$\nabla \vec{V}'_{2D} = \begin{bmatrix} 0 & 0.0515 \\ -0.4091 & 0 \end{bmatrix} \quad (23)$$

which is represent as red curves in Figure 5. Despite the spatial mean angular velocity unchanged, the shape of the typical pathline/streamline becomes flatter with larger $\dot{\theta}_{max}$ but smaller $\dot{\theta}_{min}$. The pseudo-time average angular velocity is also decreased from 0.1920 to 0.1452 mainly due to the smaller $\dot{\theta}_{min}$. A third 2D velocity gradient tensor is considered as

$$\nabla \vec{V}''_{2D} = \begin{bmatrix} 0 & 0.001 \\ -0.4596 & 0 \end{bmatrix} \quad (24)$$

Now $\dot{\theta}_{min}$ becomes very small (0.001) as presented in Figure 5 (b) (the magenta curve). The elliptic pathline become too flat to fit in the same figure with the other two pathlines in Figure 5 (a). In addition, the pseudo-time average angular velocity is substantially decreased to 0.0214, which is much smaller than its counterpart spatial average angular velocity of 0.2303. In the limiting case, $\dot{\theta}_{min} \to 0$, the pathlines will become lines parallel to the y-axis which is actually the pure shear case. The pseudo-time average will become zero because when a fluid particle rotates to the y axis, its angular velocity become zero, which means the particle will stay there thus no fully circular motion can be complete.

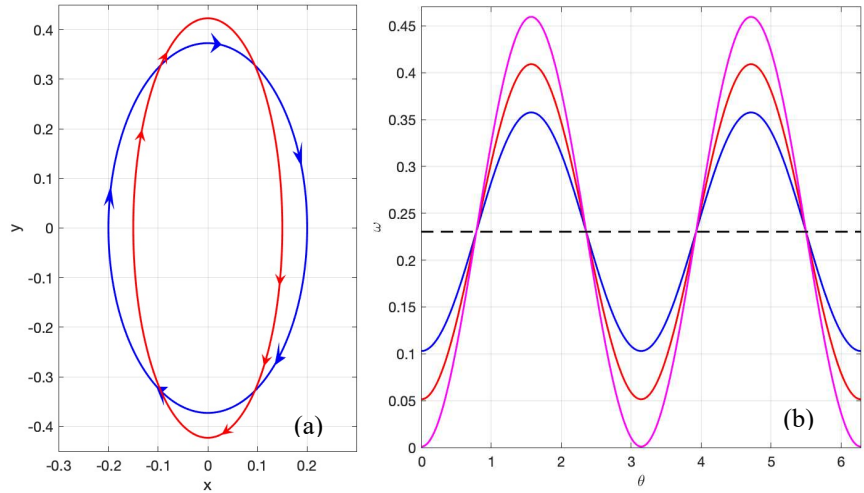

Figure 5. The induced flow by 2D velocity gradient tensor with fixed vorticity

From above discussion, even with vorticity, i.e., the spatial average angular velocity fixed, the rotational motion of the fluid can be rather diverse. Therefore, vorticity is not suitable to quantify the swirling strength. The inability of vorticity to distinguish rotational motion from background shear has been pointed out by many researchers.

Similarly, here we demonstrate that same issue could happen to $\lambda_{ci}$ by considering the 2D velocity gradient tensor with fixed $\lambda_{ci}$. The pathline induced by $\nabla \vec{V}_{2D}$ in Equation is illustrated in Figure 5 (a) and the angular velocity $\dot{\theta}$ as a function of the pseudo time $t$ in a time span of a period $T_p$ in Figure 6 by blue curves. The dashed magenta curve in Figure 6 is the corresponding sine variation with same maxim and minim for comparison while the red dashed line is spatial mean angular velocity $\frac{1}{2}(\omega_z)_P$ and the blue dashed line is $\lambda_{ci}$. With the pseudo-average angular velocity $\lambda_{ci}$ fixed, which implies the pseudo-period $T_p$ is also fixed, we consider

$$\nabla \vec{V}_{2D}''' = \begin{bmatrix} 0 & 0.0103 \\ -3.576 & 0 \end{bmatrix} \quad (25)$$

whose variation of $\dot{\theta}$ with the pseudo time $t$ is shown as the red curve in Figure 6 (b). To keep $\lambda_{ci}$ unchanged, $\dot{\theta}_{max}$ has to become very large (3.576) to compensate the low angular velocity in the region near the major axis of the elliptic trajectory. In the limiting case, $\dot{\theta}_{min} \to 0$, to keep $\lambda_{ci}$ unchanged, $\dot{\theta}_{max}$ has to become infinity. From above discussion, $\lambda_{ci}$ cannot determine the rotational motion of the surround fluid due to the fact that it is also a mean quantity like vorticity, despite one is pseudo time average while the other is spatial average.

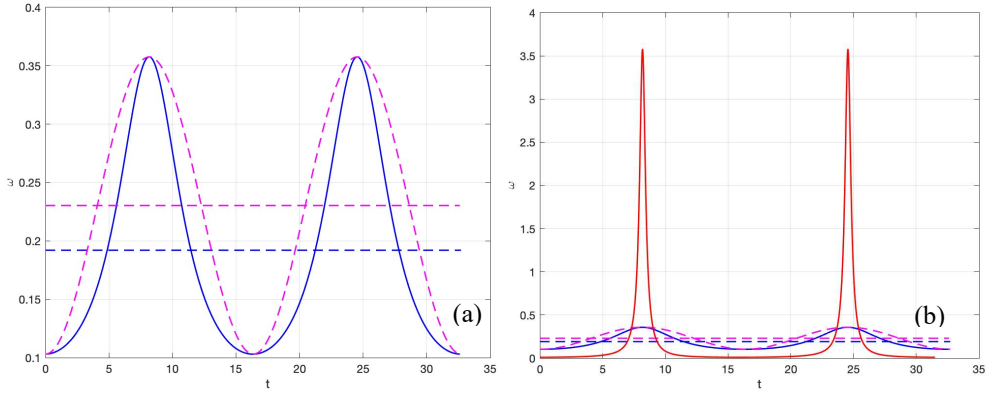

Figure 6. The induced flow by 2D velocity gradient tensor with fixed $\lambda_{ci}$

The last option is to use $\dot{\theta}_{min}$ to quantity the swirling strength which is adopted in the framework of Liutex. According to the RS decomposition proposed in Liutex, $\nabla \vec{V}_{2D}$ can be decomposed as

$$\nabla \vec{V}_{2D} = \begin{bmatrix} 0 & \dot{\theta}_{min} \\ -\dot{\theta}_{max} & 0 \end{bmatrix} = \begin{bmatrix} 0 & \dot{\theta}_{min} \\ -\dot{\theta}_{min} & 0 \end{bmatrix} + \begin{bmatrix} 0 & 0 \\ -(\dot{\theta}_{max} - \dot{\theta}_{min}) & 0 \end{bmatrix} = R + S \quad (26)$$

For the particular $\nabla \vec{V}_{2D}$ considered in Equation 22, we have

$$\nabla \vec{V}_{2D} = \begin{bmatrix} 0 & 0.1030 \\ -0.3576 & 0 \end{bmatrix} = \begin{bmatrix} 0 & 0.1030 \\ -0.1030 & 0 \end{bmatrix} + \begin{bmatrix} 0 & 0 \\ -0.2546 & 0 \end{bmatrix} \quad (27)$$

The validation of such decomposition is guaranteed by the linear nature of Equation. The resulting $R$ is the pure rotation part while $S$ is the pure shear part. It is clearer when we think backwardly to combine a pure shear and a pure rotation as shown in Figure 7. In the first row, the velocity vectors are drawn at selected points, while the streamlines of the flow are shown in the second row. The pure rotation is shown in the first column with vorticity magnitude of $2a$ ($a$ is a positive constant) in Figure 7(a) with circular trajectories shown in Figure 7 (d). The pure shear part is represented by $\partial v / \partial x = -b$, where $b$ is also a positive constant with its straight-line streamlines shown in Figure 7 (e). A combination of these pure rotation and pure shear will result in elliptic shaped streamlines as shown in Figure 7 (f) and corresponding velocity vectors on the $x$ and $y$ axes in Figure 7(c). It is thus understood that the Liutex is a systematic and mathematical way of finding the pure rotation part and the pure shear part from the velocity gradient tensor. In addition, the rotational part is uniquely determined even when the shear part changed.

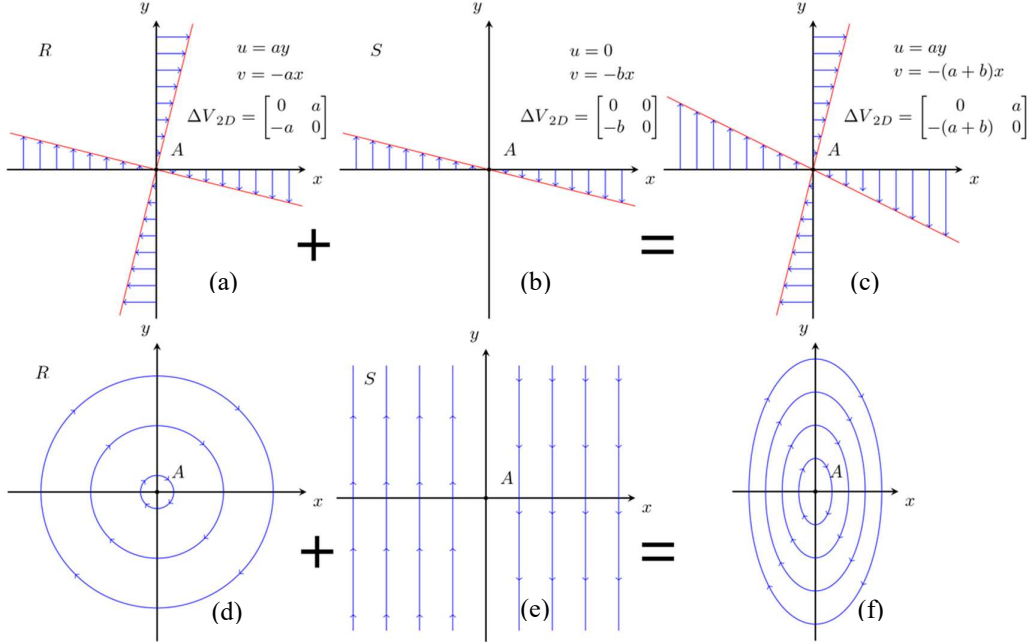

Figure 7. A sketch of RS decomposition

## 4. An explicit Liutex definition

From the above discussion, we noticed from Equation 15 that

$$\dot{\theta}_{min} + \dot{\theta}_{max} = \langle \vec{\omega}, \vec{r} \rangle \tag{28}$$

In addition, according to Equation 19, we have

$$\dot{\theta}_{min}\dot{\theta}_{max} = \lambda_{ci}^2 \tag{29}$$

Combine above two equations, we can get

$$R = 2\dot{\theta}_{min} = \langle \vec{\omega}, \vec{r} \rangle - \sqrt{\langle \vec{\omega}, \vec{r} \rangle^2 - 4\lambda_{ci}^2} \tag{30}$$

Thus, the Liutex vector can be defined as:

$$\vec{R} = R\vec{r} = \left\{ \langle \vec{\omega}, \vec{r} \rangle - \sqrt{\langle \vec{\omega}, \vec{r} \rangle^2 - 4\lambda_{ci}^2} \right\} \vec{r} \tag{31}$$

That means once we have the velocity gradient tensor, and have calculated the eigenvalue and eigenvector, we can easily get the Liutex vector according to the above equation. This is the first time such an explicit expression of Liutex vector $\vec{R}$ in terms of vorticity vector, eigenvalues and eigenvector of velocity gradient tensor is proposed instead of the two-step $Q$ and $P$ rotation as introduced in the second section. In addition, a clear physical understanding of the terms in Equation 31 can be provided. (1) $\vec{r}$ is the real eigenvector which represents the rotational axis, i.e., the direction of Liutex; (2) $\langle\vec{\omega},\vec{r}\rangle$ is the magnitude of vorticity in the direction of $\vec{r}$; (3) $\sqrt{\langle\vec{\omega},\vec{r}\rangle^2 - 4\lambda_{ci}^2}$ is the pure shear part. That means the expression itself is actually the $RS$ decomposition that vorticity can be decomposed into a rotational part $R$ and a shear part $S$.

In addition to the physical intuition provided by Equation 31, it can also help to further improve the computational efficiency. Gao and Liu remarkably decreased the computational consumption by pointing out the rotational axis is actually the real eigenvector of velocity gradient tensor. On the other hand, Equation 31 can help shorten the time needed for calculating the magnitude of Liutex vector. Here, we compare the efficiency improvement brought by Equation 31 with the version of Gao and Liu[22]. The Fortran program to compute Liutex vector consists of four parts: reading mesh and solution files ($t - read$), calculating velocity gradient tensor ($t - vg$), calculating Liutex vector ($t - L$) and writing results to files ($t - write$). The only modification to the code is in the third part: calculating Liutex vector. Table I gives the computational times of different parts of the code of previous and present methods when applied to the DNS data of a natural boundary layer transition on a flat plate, which are generated by the code DNSUTA. All these computations are done on a MacBook Pro (2017) laptop with 2.9 GHz CPU and 16 GB memory.

Table I. Comparison of computational times

|  | $t - read$ | $t - vg$ | Previous $t - L$ | Present $t - L$ | $t - write$ |
|---|---|---|---|---|---|
| Time(s) | 4.63 | 1.66 | 2.76 | 1.75 | 3.74 |

In terms of only the third part of the code, the time consumed to obtain Liutex vector from velocity gradient tensor is substantially decreased from $2.76s$ to $1.75s$ for this one time-step DNS numerical data. The increase of efficiency can be measured as $(2.76 - 1.75)/2.76 \approx 37\%$, which is rather impressive. However, if the whole procedure is considered, the efficiency increase becomes $(2.76 - 1.75)/(4.63 + 1.66 + 2.76 + 3.74) \approx 7.9\%$, which seems not very significant. We argue that the efficiency improvement provided by Equation 31 could still be very important if we integrate the RS decomposition inside Navier-Stokes equation solvers as modelling parameters of turbulence model, for example, based on its import physical meanings discussed above, which will require the solving of RS decomposition every time step.

## 5. Discussion of the RS decomposition

The Liutex method provides a logical, unique, mathematical and systematic way to identify vortices in flows especially in turbulence. In addition, the $RS$ decomposition can be a powerful tool to give insights of the dynamics of the flow. The shear part $S$ acts like an engine which provides energy for the fluid to rotate, i.e., the rotational part $R$. To illustrate this point, the method is applied to a snapshot of a natural boundary layer transition from direct numerical simulation, which is the same data we utilized in the previous section to test the computational efficiency. The numerical data are generated by code DNSUTA, which applies a sixth-order compact scheme[23] in the streamwise and normal directions while adopting the pseudo-spectral method with periodic boundary condition. An implicit sixth-order compact filter is used to eliminate spurious numerical oscillations caused by central difference schemes. The results have been validated by comparing to experiments[24] and other's DNS results[24]. For a detailed description, please refer to Liu et al.[25] and Liu and Chen[26].

In Figure 8 (a), $|R| = 0.1$ is selected as the threshold of the iso-surface to show the whole transition process, while Figure 8 (b) and (c) are enlarged views of the transitional region and the turbulent region. It can be seen that the Λ-vortices and hairpin vortices which are typical structures found in transitional flows are precisely captured with Liutex method. To visualize the distribution of the shear $S$, iso-surface of $S = 2$ is shown in Figure 9. Unlike the vortical structures, the concentration of shear $S$ is always located close to the wall from transitional region to fully turbulent region. The distance from the wall to the maximum shear level measured in wall units is roughly 10, which belongs to the buffer layer where the turbulence is most active. This shear region acts like the engine to the regeneration cycle of turbulence, and transfer to $R$ due to instability of the shear.

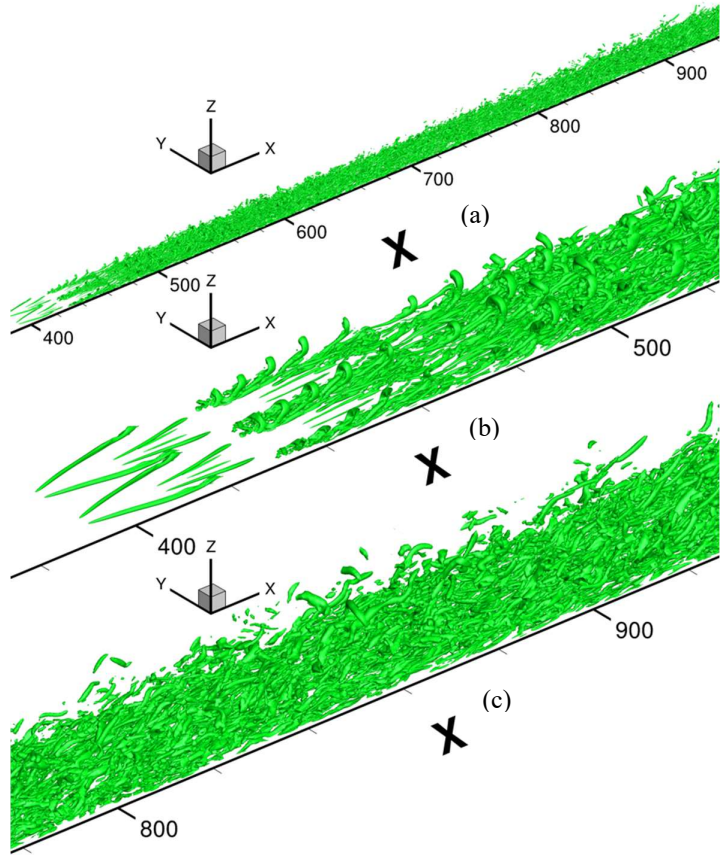

Figure 8. Liutex applied to a natural boundary layer transition flow

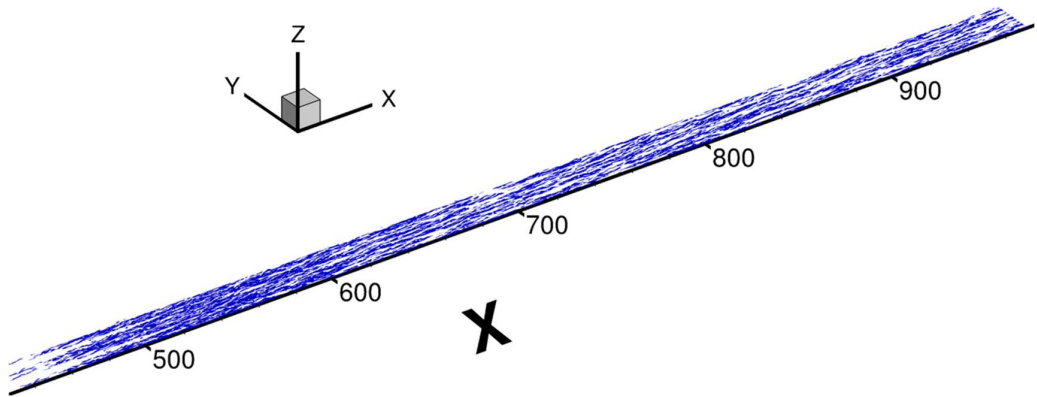

Figure 9. Iso-surface of $|S| = 2$ for the boundary layer transition

A second example of turbulent channel flow with $Re_\tau = 180$ is simulated using the open-source finite difference software 'Incompact3d'[27,28]. The dimension of the computational domain is $4\pi h \times 2h \times 2\pi h$ which discretized on a Cartesian grid of $192 \times 129 \times 128$ grid points along streamwise, normal and spanwise directions respectively, where $h$ is half the channel height. Iso-surfaces of $|R| = 2$ and $|S| = 7$ are shown in Figure 10. Similarly, concentration of $S$ is found near the wall with a distance from the wall around 10 wall units. These two examples illustrate how the RS decomposition is applied in realistic flows of turbulence, which could be helpful and informative to the understanding of the flow physics.

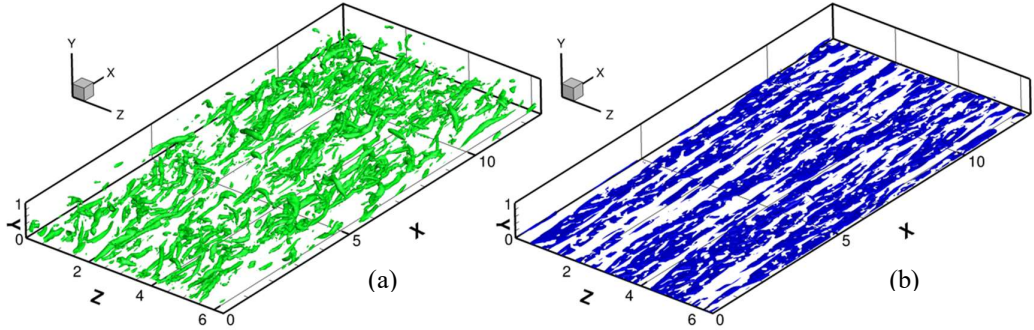

Figure 10. Liutex applied to a turbulent channel flow with $Re_\tau = 180$

## 6. Conclusions

The current study revisits the physical meaning of vorticity, $\lambda_{ci}$ and Liutex magnitude by investigate the surrounding fluid motion of a point with velocity gradient $\nabla \vec{V}$. It is shown that the vorticity along the direction of rotational axis is the spatial average angular velocity while $\lambda_{ci}$ is pseudo-time average angular velocity. Then a discussion of why these two quantities cannot solely determine the rotational motion is given, and the logic to use the minimum angular velocity as proposed by Liutex is elaborated. Based on these understanding, an explicit expression of Liutex vector in terms of vorticity, eigenvalues and eigenvectors of velocity gradient tensor is given for the first time. The physical intuition and computational efficiency improvement provided by this expression is then discussed. Finally, the importance of RS decomposition is reiterated with two examples of boundary layer transition and turbulent channel flow.

## Acknowledgments

Dr. Yiqian Wang is partly supported by the National Natural Science Foundation of China (No. 11702159) and China Post-Doctoral Science Foundation (No. 2017M610876). This work is accomplished by using code DNSUTA released by Dr. Chaoqun Liu at the University of Texas at Arlington in 2009.

# References

[1] J.-Z. Wu, H.-Y. Ma, and M.-D. Zhou, Vorticity and Vortices Dynamics (Springer-Verlag, Berlin Heidelberg, 2006).

[2] X. Dong, G. Dong, and C. Liu, "Study on vorticity structures in late flow transition," Phys. Fluids **30**, 014105 (2018).

[3] Y. Wang, Y. Yang, G. Yang, and C. Liu, "DNS study on vortex and vorticity in late boundary layer transition," Commun. Comput. Phys. **22**, 441-459 (2017).

[4] S. K. Robinson, "Coherent motion in the turbulent boundary layer," Annu. Rev. Fluid Mech. **23**, 601–639 (1991).

[5] J. Jiménez, "Coherent structures in wall-bounded turbulent," J. Fluid Mech. **842**, P1 (2018).

[6] T. Theodorsen, "Mechanism of turbulence," in Proceedings of the Midwestern Conference on Fluid Mechanics (Ohio State University, Columbus, OH, 1952).

[7] R. J. Adrian, "Hairpin vortex organization in wall turbulence," Phys. Fluids **19**, 041301 (2007).

[8] Y. Wang, H. Al-Dujaly, Y. Yan, N. Zhao and C. Liu, "Physics of multiple level hairpin vortex structures in turbulence", Sci. China: Phys., Mech. Astron. **59**, 624703 (2016).

[9] G. Eitel-Amor, R. örlú, P. Schlatter, and O. Flores, "Hairpin vortices in turbulent boundary layers," Phys. Fluids **27**, 025108 (2015).

[10] S. J. Kline, W. C. Reynolds, F. A. Schraub and P. W. Runstadler, "The structure of turbulent boundary layers," J. Fluid Mech. **30**, 741-773 (1967).

[11] W. Schoppa and F. Hussain, "Coherent structure generation in near-wall turbulence," J. Fluid Mech. **453**, 57-108 (2002).

[12] M. Chong, A. Perry, and B. Cantwell, "A general classification of three dimensional flow fields," Phys. Fluids A **2**, 765–777 (1990).

[13] J. Hunt, A. Wray, and P. Moin, "Eddies, streams, and convergence zones in turbulent flows," Report CTR-S88, Center For Turbulence Research, 1988.

[14] J. Jeong and F. Hussain, "On the identification of a vortices," J. Fluid Mech. **285**, 69–94 (1995).

[15] C. Liu, Y. Wang, Y. Yang, and Z. Duan, "New omega vortex identification method," Sci. China: Phys., Mech. Astron. **59**, 684711 (2016).

[16] H. Chen, R. J. Adrian, Q. Zhong, and X. Wang, "Analytic solutions for three dimensional swirling strength in compressible and incompressible flows," Phys. Fluids **26**, 081701 (2014).

[17] Q. Chen, Q. Zhong, M. Qi, and X. Wang, "Comparison of vortex identification criteria for velocity fields in wall turbulence," Phys. Fluids **27**, 085101 (2015).

[18] Y. Zhang, X. Qiu, F. Chen, K. Liu, X. Dong, and C. Liu, "A selected review of vortex identification methods with applications," J. Hydrodyn. **30**(5), 767-779 (2018).

[19] B. Epps, "Review of vortex identification methods," AIAA paper 2017-0989, 2017.

[20] C. Liu, Y. Gao, S. Tian, and X. Dong, "Rortex—A new vortex vector definition and vorticity tensor and vector decompositions," Phys. Fluids **30**, 035103 (2018).

[21] S. Tian, Y. Gao, X. Dong, and C. Liu, "Definition of vortex vector and vortex," J. Fluid Mech. **849**, 312–339 (2018).

[22] Y. Gao and C. Liu, "Rortex and comparison with eigenvalue-based vortex identification criteria," Phys. Fluids **30**, 085107 (2018).

[23]S.K. Lele, "Compact finite difference schemes with spectral-like resolution," J. Comput. Phys. **103**, 16-42 (1992).

[24]C. Lee and R. Li, "Dominant structure for turbulent production in a transitional boundary layer," J. Turbul. **8**, 55 (2007).

[25]C. Liu, Y. Yan, and P. Lu, "Physics of turbulence generation and sustenance in a boundary layer," Comput. Fluids **102**, 353–384 (2014).

[26]C. Liu and L. Chen, "Parallel DNS for vortex structure of late stages of flow transition," Comput. Fluids **45**, 129-137 (2011).

[27]S. Laizet and E. Lamballais, "High-order compact schemes for incompressible flows" A simple and efficient method with quasi-spectral accuracy," J. Comput. Phys. **228**, 5989-6015 (2009).

[28]S. Laizet and N. Li, "Incompact3d: A powerful tool to tackle turbulence problems with up to O(105) computational cores," Int. J. Numer. Meth. Fluids **67**, 1735-1757 (2011).